# Models & Metaphors
## Part II: The Aesthetical Conception of Natural Laws

**Rainer E. Zimmermann[1], Annette Schlemm, Doris Zeilinger,**
IAG Philosophische Grundlagenprobleme,
Fachbereich 1, Universität,
Nora-Platiel-Str. 1, D – 34127 Kassel
e-mail: pd00108@mail.lrz-muenchen.de

**Vladimir G. Budanov, Vjacheslav E. Voitsekhovitch**
Institute of Philosophy,
Department of Interdisciplinary Studies,
Russian Academy of Sciences,
14 Volkhonka, R – 119842 Moscow
e-mail: nsavicheva@mtu-net.ru

### Abstract

Following the results and the general methodological framework of a preceeding paper (Models & Metaphors, part I, henceforth: MM), we develop some results on the epistemological state of natural laws and ask for explicit possibilities of mediating the contexts of nature (in terms of the sciences) and aesthetics.

### 1. Origin and State of Natural Laws

The topic of laws has been a characteristic field of the philosophy of science proper rather than of a philosophy of nature of classical type. Hence, primarily, it is the analytical schools such as those of Stegmueller [1] who have dealt with this in some detail so far. There is also a certain tradition of research on this topic in the former GDR (partly suppressed there at the time) which has not yet been received very prominently up to now. [2] However, if visualizing science as a specific form of social activity, the original topic is introduced into a wider scope of problems. As we have seen in [MM], the basic question is for the fundamental social relationships which are at the roots of defining laws at all. We have found these relationships with a view to the original mediation of perception, communication, and design. In other words: What we expect to find now is a panoramatic view onto the various forms of mediation among different fields of approach of which the most important are psychological and sociological aspects of the process of scientific production (otherwise called *history of science*

---

[1] Also: Clare Hall, UK-Cambridge CB3 9AL. Present Address: Lehrgebiet Philosophie, FB 13 AW, FH, Lothstr.34, D – 80335 Muenchen

in the sense of Foucault, Serres and others [3]). The point is simply that given a universal constitution of the scientific orientation within the world, and also given some more insight into the mediation of perception, communication, and design (in anthropological terms), it is primarily the social process which actually models (in the outcome) the results of scientific insight. On the other hand, a differentiation is necessary between scientific concepts of law and everyday concepts of law such as those coined in legal terms. Obviously, the one cannot be disconnected from the other, and the everyday utilization of the concept of law is more than just a formal metaphor: If modeling turns out as an implicity anthropologically organized human activity rather than an objective mode of extracting knowledge from something which there really is in absolute terms, then laws are not only human-made, but they are themselves part of the modeling procedure. And this is also true for the laws of nature then. Hence, there is an obvious relationship between the legal aspects of law (as means of regulating subjective human behaviour) on the one hand, and „global" aspects of laws (of nature) as means of regulating objective behaviour of (non-human) matter. [4] Hence, the origin of laws would be human at any rate, while the state of the (view of) laws would depend on the explicit evolution of the social system rather than on the modeled evolution of the world. Unless, the one is being visualized as a special aspect of the other. And this would be indeed Schelling's view.

## 2. The Mediation of Nature and Aesthetics

As a modern (and generalized) version of Schelling's philosophy, the philosophy of Ernst Bloch can be visualized as one which offers both an open perspective as to the first topic mentioned here and to the mediation of nature and aesthetics on the other hand. The Blochian starting point is the existential enigma which characterizes an unfolding world whose ontological state would be one of the *not-yet* (which is still becoming) rather than something which has already come out of its original field of possibilities (and is actually being). [5] This idea refers implicitly to what Hogrebe recognizes in Schelling's *world formula* when characterizing the irreducible and immanent self-inconsistency of the world's initial singularity, at the same time not only a dynamical *propelling-forward* of the process, but also an epistemic exaction. [6] According to Hogrebe, the latter is in fact nothing but the origin of aesthetics, in so far as this underlying inconsistency asks for re-founding of a foundation which is nothing but non-foundation (a genuine Schellingian concept). Consequently, for his entry into nature, Bloch does neither look for the traditional sciences, nor for a praxis which is rooted in ideology and exploitation. In a sense, Bloch refers not only to Schelling, but also to Spinoza when alluding to a nature which is self-explicative in an ontological as well as epistemological double-meaning. Aesthetics visualized as a direct consequence of the epistemic problem of a dynamically unfolding matter of the world takes the position of a „bridge between nature and humans". On the one hand, this carries the connotation of separating the one from the other (in epistemological terms), on the other hand, it carries the connotation of implying that within the aesthetical modeling of nature, there is an intrinsic *shining-forth* of something which has not yet come into being, but which eventually may become in due time. Bloch himself gives a large number of examples for this mediating function of the arts. [7] In terms of what we are discussing here, this conception can be visualized with a view to the central role of language as an approach which may be able to actually translate the systematic concepts developed in [MM] into everyday details of the worldly process (or into its „traces", to utilize a Blochian expression).

### 3. Conclusions

What we have done here is to actually put forward preliminary results as to the further development of epistemic criteria for what we have called *human strategies in complexity*. Obviously, this conception includes both research dealing with the basic pre-conditions of the interactions among perception, communication, and design on the one hand (defining the fundamental settings of anthropologically consistent ways of grasping the world), and the immediate applications within the field of the explicit modeling in practise. This is basically a strictly interdisciplinary (and in fact ancient) task which relates ontology to epistemology, and to ethics as well. [8]

### 4. Acknowledgements

The authors would like to thank the INTAS/NIS cooperation organization of the European Commission at Brussels for financial support under contract number MP/CA 2000-298. We thank also the other members in the various subtasks of this cooperation, in particular Wolfgang Hofkirchner of the University of Technology, Vienna, as chief organizer.